# Silicon Resonant Antenna

Ali Ayazi, and Bahram Jalali

Department of Electrical Engineering, University of California, Los Angeles, CA 90095-1594

*Abstract* — **The need for high data rate communication systems has been dramatically increasing in recent years. The millimeter-wave bands (such as 60 GHz) have been investigated for wireless high data rate transmission. At these frequencies, the conductor loss is significant which makes dielectric resonator antenna (DRA) a good replacement for microstrip patch antennas currently in use. Moreover, use of silicon as the dielectric material in a DRA provides the great feature of bandwidth tunability through carrier density modulation. In this paper, we present experimental and theoretical studies of antennas made from silicon.**

## I. INTRODUCTION

Millimeter-wave bands have attracted a lot of attention in recent years. For applications today, frequencies beyond the usual microwave band and in the millimeter region are often required. One of the main motivations is the availability of multi-GHz spectrum. For example 60 GHz band is considered for short range indoor broadband communications because oxygen absorption and obstacles at the proximity of 60GHz introduce strong attenuation which will provide isolation from nearby transmitters [1,2].

Microstrip patch elements [3] and arrays [4] have been successfully implemented at millimeter-wave frequencies. However, at millimeter-wave frequencies, conductor loss is significant [5]. Dielectric resonator antennas (DRA's) with no inherent conductor loss are efficient antenna elements at these frequencies. DRA also has the inherent advantage of a larger bandwidth over a microstrip patch antenna [6].

Since dielectric loss is the only source of loss inside the DRA, modifying loss tangent (tan δ) of the material tunes the bandwidth. In a semiconductor, changing the carrier concentration modifies the conductivity of the material and hence changing loss tangent (tan δ). This immediately suggests that by using a semiconductor as the dielectric material of the DRA, it is possible to tune the bandwidth of the antenna by somehow controlling the carrier concentration. Optical excitation and temperature change are two possible ways to control the carrier concentration.

In this paper, we propose the use of silicon as antenna material and report theoretical design of a 60 GHz antenna plus experimental results on a 7.5 GHz device.

## II. DEPENDENCE OF LOSS TANGENT ON CARRIER CONCENTRATION IN SILICON

According to Maxwell's equations,

$$\nabla \times H = j\omega \left[ \varepsilon' + \left( \varepsilon'' + \sigma/\omega \right) \right] E \qquad (1)$$

Where $\varepsilon'' + \sigma/\omega$ is the effective imaginary part of the permittivity. The loss tangent of a dielectric medium is defined as the ratio of the imaginary part of the permittivity to its real part. Then we have,

$$\tan \delta_T = \frac{\omega \cdot \varepsilon'' + \sigma}{\omega \cdot \varepsilon'} = \frac{\varepsilon''}{\varepsilon'} + \frac{\sigma}{\omega \cdot \varepsilon'} = \tan \delta_D + \tan \delta_L \qquad (2)$$

Where $\varepsilon''/\varepsilon'$ is the intrinsic loss (tan $\delta_D$) resulting from the polarization loss of the intrinsic silicon and $\sigma/\omega \cdot \varepsilon'$ is the extrinsic loss (tan $\delta_L$) due to the finite conductivity of silicon which is coming from free electron and holes [7]. So,

$$\tan \delta_L = \frac{\sigma}{\omega \cdot \varepsilon'} = \frac{q\left( \mu_n \cdot \Delta N_n + \mu_p \cdot \Delta N_p \right)}{2\pi f \cdot \varepsilon_r' \cdot \varepsilon_o} \qquad (3)$$

Assuming an intrinsic loss of tan $\delta_D$ = 0.001 at 60 GHz [8,9], we get to the curve in Fig. 1 for the total loss tangent of silicon vs. free carrier concentration.

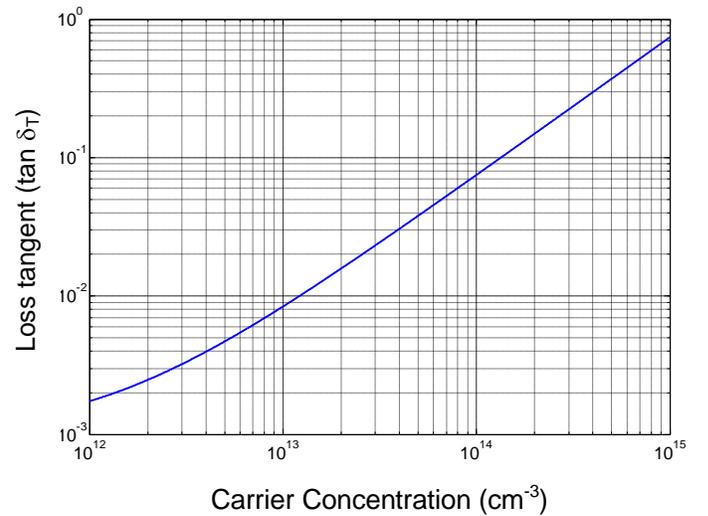

Fig. 1. Calculated loss tangent of silicon vs. carrier concentration. Electron and hole excess carrier concentrations are assumed to be equal. $\mu_n$ = 1350 cm$^2$/V.s, $\mu_p$ = 480 cm$^2$/V.s, and $f$ = 60 GHz.

This curve suggests that by introducing free carriers into silicon DRA through, it is possible to increase loss tangent by several orders of magnitude. At the same time, tuning loss tangent will modify bandwidth.

### III. SILICON DRA FOR C BAND

First, as a proof of concept, a silicon DRA is fabricated to work at C band (4-8 GHz). It is a cylindrical DRA with $D$ = 25mm and $h$ = 19.35mm. Figure 2 shows the result of HFSS simulation of this structure.

It has been excited by an electric field probe which is suitable to excite TM modes in a cylindrical DRA. Two resonant peaks in $S_{11}$ curve can be seen. Electric and magnetic field profiles confirm that these two modes are associated with $TM_{011+\delta}$ and $TM_{012+\delta}$ respectively.

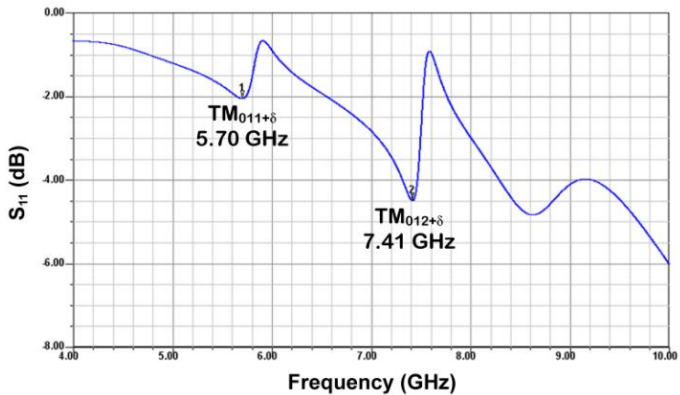

Fig. 2. Simulation result for cylindrical DRA made of silicon. Two TM resonant modes observed.

Next, $S_{11}$ is measured on the silicon sample. It is observed in Fig. 3 that good agreement is obtained between simulation from HFSS modeling and measurement of $S_{11}$ which clearly verifies the existence of these two modes in practice.

### IV. SILICON DRA AT 60 GHZ

The dimensions of the DRA scales with the free space wavelength so as the frequency goes up, DRA becomes smaller.

Using the dielectric waveguide model described by Mongia et al. [10] the dimensions of a rectangular DRA to have a $TE^z_{111}$ mode at 60 GHz can be determined. Aperture coupling through microstrip slot is selected to excite the DRA [11]. Excitation schematic can be seen in Fig 4.

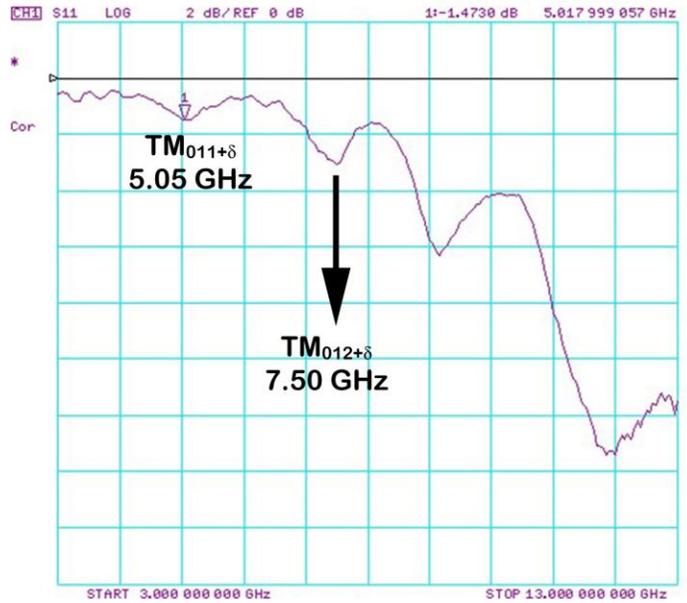

Fig. 3. Result of the measured $S_{11}$ verifying the existence of the same two TM modes seen in simulation of Fig. 2. (2dB/unit for vertical axis)

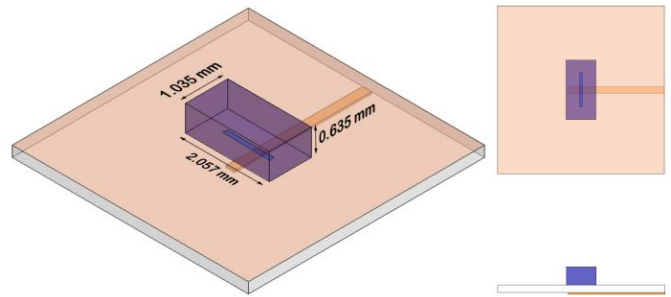

Fig. 4. 3D schematic of the rectangular aperture coupled DRA to work at 60 GHz.

HFSS simulation in Fig. 5 confirms this mode and shows a clear resonance at 61.60 GHz (Fig. 5).

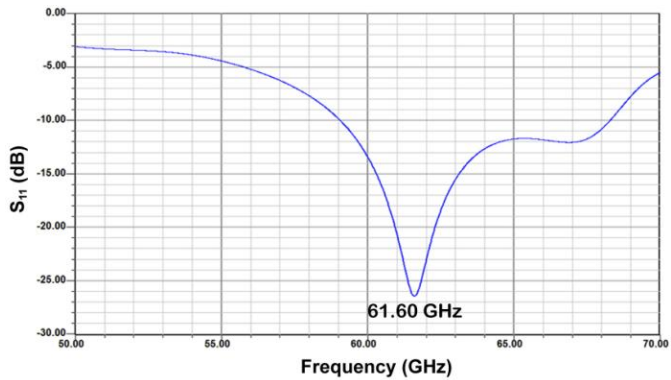

Fig. 5. $S_{11}$ simulation result of the aperture coupled DRA of Fig. 4.

IV. CONCLUSION

The results obtained have demonstrated that a dielectric resonator antenna from silicon can be used as an antenna element for 60 GHz band. Its fabrication is considerably easier than similar antennas at this frequency range using microstrip patch antennas. Moreover, it provides the great feature of controlling the bandwidth through modulation of carrier density.